# Electrical 180º switching of Néel vector in spin-splitting antiferromagnet

**Authors:**
Lei Han[1]†, Xizhi Fu[2]†, Rui Peng[2], Xingkai Cheng[2], Jiankun Dai[1], Liangyang Liu[3], Yidian Li[3], Yichi Zhang[1], Wenxuan Zhu[1], Hua Bai[1], Yongjian Zhou[1], Shixuan Liang[1], Chong Chen[1], Qian Wang[1], Xianzhe Chen[1], Luyi Yang[3,4,5], Yang Zhang[6,7], Cheng Song[1]*, Junwei Liu[2]*, and Feng Pan[1]*

**Affiliations**
[1] Key Laboratory of Advanced Materials (MOE), School of Materials Science and Engineering, Tsinghua University, Beijing 100084, China.
[2] Department of Physics, The Hong Kong University of Science and Technology, Hong Kong 999077, China.
[3] State Key Laboratory of Low Dimensional Quantum Physics, Department of Physics, Tsinghua University, Beijing 100084, China.
[4] Frontier Science Center for Quantum Information, Beijing 100084, China.
[5] Collaborative Innovation Center of Quantum Matter, Beijing 100084, China.
[6] Department of Physics and Astronomy, University of Tennessee, Knoxville, Tennessee 37996, USA.
[7] Min H. Kao Department of Electrical Engineering and Computer Science, University of Tennessee, Knoxville, Tennessee 37996, USA.
†These authors contributed equally to this work.
*Corresponding author. Email: songcheng@mail.tsinghua.edu.cn; liuj@ust.hk; panf@mail.tsinghua.edu.cn.

**Abstract**
Antiferromagnetic spintronics have attracted wide attention due to its great potential in constructing ultra-dense and ultra-fast antiferromagnetic memory that suits modern high-performance information technology. The electrical 180º switching of Néel vector is a long-term goal for developing electrical-controllable antiferromagnetic memory with opposite Néel vectors as binary "0" and "1". However, the state-of-art antiferromagnetic switching mechanisms have long been limited for 90º or 120º switching of Néel vector, which unavoidably require multiple writing channels that contradicts ultra-dense integration. Here, we propose a deterministic switching mechanism based on spin-orbit torque with asymmetric energy barrier, and experimentally achieve electrical 180º switching of spin-splitting antiferromagnet $Mn_5Si_3$. Such a 180º switching is read out by the Néel vector-induced anomalous Hall effect. Based on our writing and readout methods, we fabricate an antiferromagnet device with electrical-controllable high and low resistance states that accomplishes robust write and read cycles. Besides fundamental advance, our work promotes practical spin-splitting antiferromagnetic devices based on spin-splitting antiferromagnet.

**Teaser**
The electrical 180º switching and detection of the Néel vector in spin-splitting antiferromagnet $Mn_5Si_3$ is achieved.

# MAIN TEXT
## Introduction

One of the most well-known applications of spintronics is the electrical-controllable non-volatile magnetic random-access memory (MRAM), where the information states "1" and "0" are stored by the up and down directions of the order parameter in ferromagnet (FM), i.e., the magnetic moment (*1-3*). For the original MRAM, the electrical read is based on tunneling magnetoresistance (*4-6*) (TMR), and the electrical write is realized by applying current along single channel to switch the magnetic moment due to spin-transfer torques (*7, 8*) (STTs). Over the past decades, although many emerging effects such spin-orbit torques (SOTs) switching (*9, 10*) have improved the performance of MRAM, intrinsic large stray fields and GHz dynamics of FM fundamentally limit MRAM from achieving future breakthroughs in high-density integration and high-speed operation (*11, 12*). Besides, a very small magnetic field may be able to switch the direction of magnetic moment in MRAM, challenging its data storage stability (*11, 12*). Supreme to FM counterparts, antiferromagnet (AFM) exhibits tremendous potential for building ultra-dense and ultra-fast AFM-RAM with high immunity to magnetic field disturbance thanks to their intrinsic advantages of zero stray field, terahertz dynamics, and compensated moment (*13-17*). With TMR predicted (*18, 19*) and discovered (*20, 21*) recently in many AFMs, the long-desired electrical-controllable AFM-RAM becomes possible, as long as the electrical switching of the AFM order parameter (i.e., the Néel vector) can be achieved.

However, the state-of-art reorientation of Néel vectors are all in-plane 90º or 120º switching by SOTs, which inevitably require multiple writing channels, leading to huge integration difficulties for AFM-RAM (*13, 22-24*). Similar to FM MRAM, practical AFM-RAM undoubtably needs electrical current along single writing channel to switch the Néel vector back and forth, which generates high and low resistance to represent binary "1" and "0" (*11*). An ideal solution is the 180º switching of Néel vectors back and forth ($n_+ \leftrightarrow n_-$) in AFMs by electrical current with opposite polarities along single writing channel. Moreover, 180º switching is truly needed to generate substantial antiferromagnetic TMR (*18, 19*). Therefore, electrical 180º switching of Néel vector is very important, but remains unveiled.

The core difficulty of electrical 180º switching of Néel vector lies in equal energy barriers for $n_+ \rightarrow n_-$ and $n_- \rightarrow n_+$, defined by uniaxial magnetic anisotropy. In this case, if SOTs drive the Néel vector to rotate from $n_+$ to $n_-$ by overcoming the energy barrier for $n_+ \rightarrow n_-$, it can also continuously drive the Néel vector to rotate from $n_-$ to $n_+$ (*23*), which makes the 180º switching of Néel vector non-deterministic. That is why the more promising 180º switching of Néel vector remains unrealized for a long time since the first discovery of 90º switching of Néel vector in 2016 (*13*). Note that the 180º switching of CuMnAs by staggered spin-orbit fields is actually due to antiferromagnetic domain wall motion instead of the most desired deterministic 180º switching of Néel vector (*25*). Besides, the 180º switching of perpendicular cluster magnetic octuple has been achieved in non-collinear AFM $Mn_3Sn$, but it is fundamentally different from 180º switching of Néel vector as the order parameter in collinear AFM (*26, 27*). Another important difficulty is to readout the 180º switching of Néel vector. Antiferromagnetic anisotropic magnetoresistance has been widely adopted but it cannot read out 180º switching (*13, 22-24*). Harmonic measurements need specific sublattice symmetry breaking, which is not general and is hard to implement (*25*). Fortunately, the 180º switching of Néel vector may be feasibly read out by the anomalous Hall effect (*28-31*) (AHE) in some recently discovered special collinear AFMs with spin-splitting band structure (*32-37*), also termed as altermagnets (*38, 39*). Moreover, their spin-splitting band structure and related properties, such as unconventional spin current generation (*40-42*) and piezomagnetism (*37*), can all be manipulated through 180º switching of the Néel vector, which makes it very promising to investigate the electrical 180º switching of Néel vector in spin-splitting AFM.

Herein, we achieve deterministic electrical 180º switching of the Néel vector in spin-splitting AFM $Mn_5Si_3$ films, based on spin-orbit torque with asymmetric switching barrier that can be easily generalized to other AFM materials. We prove theoretically and experimentally that 180º switching of the Néel vector indeed tunes the spin-splitting bands and flips the anomalous Hall conductivity (AHC), which can be used as an unconventional electrical readout approach of 180º switching. With these write and read methods ready, we successfully fabricate an AFM device with high and low resistance that realizes robust write and readout cycles, paving the way for the long-desired electrical-controllable AFM-RAM.

## Results

### Spin splitting manipulated by reversing the Néel vector

We start with introducing basic crystal structure and magnetic phase characteristics of $Mn_5Si_3$. The space group of $Mn_5Si_3$ at room temperature is $P6_3/mcm$ with the unit cell composing of 4 $Mn_a$ and 6 $Mn_b$ atoms at two inequivalent Wyckoff positions and 6 Si atoms (Fig. 1A) (*43-45*). As temperature decreases, $Mn_5Si_3$ undergoes a magnetic phase transition from paramagnetic to collinear AFM (cAFM), and then to non-collinear AFM (ncAFM). In the cAFM phase, the G-type AFM ordering occurs on two-thirds of $Mn_b$ atoms, where adjacent magnetic $Mn_b$ atoms have opposite spin orientations to form the Néel vector (Fig. 1A, fig. S1 for other perspectives) (*43-47*). Clearly, the *PT* symmetry is broken in cAFM $Mn_5Si_3$ thin film (*29*), and hence its energy state at a generic momentum is spin polarized, i.e., $E_k^\uparrow \neq E_k^\downarrow$, due to the exchange coupling between itinerant electrons and local magnetic moments. Moreover, the states at mirror symmetry-related momenta must possess contrasting spin polarizations to form the *C*-paired spin-momentum locking, further suppressing spin-flipping (*37*), which is confirmed by our first-principles calculations as shown in Fig. 1B. These unique features provide possibilities for the existence of AHE in cAFM $Mn_5Si_3$ thin film (*29*).

Next, it is demonstrated that through reversing the Néel vector, we can manipulate spin-resolved and Berry curvature-dependent phenomena, which in turn serve as potential readout approach of 180º Néel vector switching in cAFM $Mn_5Si_3$. Although spin-orbit coupling has negligible effect on the band structure (fig. S1), it can break the strict SU (2) symmetry and hence induce non-zero Berry curvature for spin-splitting bands of cAFM $Mn_5Si_3$ with *C*-paired spin-momentum locking as presented in Fig. 1C. When the Néel vector *n* is switched 180º from $n_+$ to $n_-$, equivalent to the time-reversal operation, the spin orientations will be reversed (Fig. 1A). Accordingly, both spin splitting (Fig. 1B) and Berry curvature (Fig. 1C) at $(k_x, k_y)$ will switch their signs at the *T*-paired momentum $(–k_x, –k_y)$, as displayed in Fig. 1D and Fig. 1E, respectively. Therefore, through 180º Néel vector switching, we can control the spin-splitting bands and Berry curvature, and hence manipulate all spin-resolved and Berry curvature-dependent phenomena such as AHE. Correspondingly, AHE may serve as a potential readout method of 180º Néel vector switching.

### Demonstration of anomalous Hall effect as a readout method

As expected, AHE is observed distinctly in the cAFM temperature range (60–230 K) of sputtered $Mn_5Si_3$(0001) thin films (fig. S2, Supplementary text I) by Hall resistivity measurements (fig. S3) and polar magneto-optical Kerr effect measurements (fig. S4).

Typical hysteresis of Hall resistivity $\rho_{yx}$ under out-of-plane magnetic field at 150 K is shown in Fig. 2A, exhibiting a non-volatile characteristic. Notably, the corresponding hysteresis of magnetization $M$ in Fig. 2B exhibits two abrupt changes around zero field and 4 kOe: The one around zero field (black line) is due to the net moment from unavoidable local defects in the sputtering process, which does not contribute to $\rho_{yx}$ near zero field and can be drastically suppressed (fig. S5, Supplementary text II); The other around 4 kOe (blue line) has consistent coercive field $H_c$ with that of the $\rho_{yx}$ hysteresis, which can be more clearly illustrated by similar peak locations in field-derivative hysteresis of both $\rho_{yx}$ and $M$ around 4 kOe (fig. S3, Supplementary text I). The net moment around 4 kOe, denoted to be $m$, most likely originates from the relativistic Dzyaloshinskii-Moriya interaction (DMI) due to interfacial symmetry breaking (*48*).

A natural question is that whether the AHE is simply due to this net moment $m$. The answer is that AHE comes from Néel vector-dependent Berry curvature instead of $m$. First, the measured $m$ can be as small as ~1 emu cm$^{-3}$, corresponding to only 2.5 milli $\mu_B$ per magnetic Mn$_b$ atom, which can hardly induce the measured large $\rho_{yx}$ of 0.17 microhm cm with comparable magnitude to typical ferromagnets (*49, 50*). In fact, $m$ varies for samples of different thicknesses, but $\rho_{yx}$ almost remains unchanged, which further rules out that $\rho_{yx}$ comes from $m$ (Fig. 2C, fig. S6, Supplementary text II). Moreover, the coercive field of AHE drastically decreases when the width of the Hall bar is reduced below 500 nm, which is quite different from ferromagnets (fig. S7, Supplementary text II). Besides, $\rho_{yx}$ has opposite temperature dependence compared to longitudinal resistivity $\rho_{xx}$ (inset of fig. S3), which excludes out extrinsic mechanisms and indicates intrinsic mechanism (*49, 50*). Hence, the experimentally observed AHE should be attributed to the intrinsic Néel vector-dependent Berry curvature as indicated by first-principles calculations in Fig. 1. Note that there is no detectable net moment $m$ in a previous result on cAFM Mn$_5$Si$_3$ thin film, which may be due to different deposition methods and substrates (*29*). It is worth emphasizing that it is impossible to measure AHE with absolutely zero net moment, because both Hall vector and magnetization follow the same symmetry rules (*31, 49, 50*).

We further performed a systematic theoretical study through atomic spin simulations and first-principles calculations (Materials and Methods) to explicitly explain why the AHE indeed arises from $n$-dependent Berry curvature instead of the negligible net moment $m$. We first demonstrate that the 180º switching of $m$ by reversing magnetic field will simultaneously drive the Néel order $n$ to reverse, which provides prerequisites for $n$-dependent AHE measured by sweeping magnetic field. It is investigated from the responses of $m$ and $n$ to external magnetic field by simulations, considering the realistic crystal and magnetic structure of cAFM Mn$_5$Si$_3$ with all of the parameters extracted from first-principles calculations (fig. S8, S9, Supplementary text III). The simulated magnetic hysteresis is shown in Fig. 2D, where the coercive field and the magnitude of simulated $m$ match very well with experimental $m$ in Fig. 2B (Supplementary text III). Figure 2E presents the simulated trajectories of both $m$ and $n$ when sweeping the out-of-plane magnetic field $H$ from 15 kOe to –15 kOe, where the 180º switching of $m$ from $m_+$ to $m_-$ by $H$ drives the 180º switching of $n$ from $n_+$ to $n_-$ simultaneously. This process only requires overcoming the magnetic anisotropy energy (MAE) since $m$ and $n$ remain the same chirality with unchanged DMI energy. In fact, the reverse of $m$ can also be achieved without reversing $n$ by only switching the tilting direction of cAFM moment, which does not overcome the MAE but changes the chirality between $m$ and $n$, and thus increases DMI energy. However, this scenario can be excluded for cAFM Mn$_5$Si$_3$ because simply reversing $m$ without reversing $n$ cannot bring about the measured AHE, as we discussed in the previous paragraph. In addition, first-principles calculations also reveal that MAE is one order of magnitude smaller than DMI energy (Supplementary text III), further illustrating that the reverse of $m$ and $n$ will appear simultaneously in cAFM $Mn_5Si_3$.

With $m$ and $n$ reverse at the same time demonstrated, we move on to demonstrate that it is the reverse of $n$ instead of $m$ that determines the AHE by first-principles calculations. The dependence of AHC $\sigma_{xy}$ on both $m$ and $n$ in cAFM $Mn_5Si_3$ is calculated at 150 K, as shown in Fig. 2F. $\sigma_{xy}$ at the Fermi level is calculated to be approximately 400 S m$^{-1}$, matching the experimentally-obtained $\sigma_{xy}$ of around 300 S m$^{-1}$. Notably, $\sigma_{xy}$ values for $n_+$ with $m_+$ or $m_-$ are almost the same, but both are opposite to those for $n_-$ with $m_+$ or $m_-$. Thus, the influence of $m$ on $\sigma_{xy}$ is negligible, and the sign reversal of $\sigma_{xy}$ is determined by the 180º switching of $n$, consistent with aforementioned experimental results. Moreover, $\sigma_{xy}$ at the Fermi level for $n$ along any direction is calculated and presented as a spherical image in Fig. 2G. It turns out that $\sigma_{xy}$ always reverses once $n$ is 180º-switched, irrespective of any specific direction $n$ aligning. As a result, both our experimental and theoretical results reveal that the sign of $\sigma_{xy}$ is uniquely determined by the direction of $n$, and thus the AHE can serve as a reliable readout method for 180º switching of $n$ in cAFM $Mn_5Si_3$.

**Asymmetric energy barrier mechanism for electrical 180º switching of Néel vector**

After solving the problem of read the 180º switching of Néel vector, the remaining question is how to realize the deterministic electrical 180º switching of Néel vector, i.e., the information write process. There are three basic requirements in general: (1) Both $n_+$ and $n_-$ should be stable with minimum energy to realize binary information states "0" and "1"; (2) $n$ needs to be driven into motion electrically; (3) Transition from $n_+$ to $n_-$ and its time-reversal counterpart from $n_-$ to $n_+$ must have unequal energy barriers to achieve the deterministic switching. The requirement (1) is naturally satisfied in cAFMs since both $n_+$ and $n_-$ are stable along the easy axis due to MAE, and thus we will focus on how to fulfill requirements (2) and (3) in the following.

According to the definition of $m$ and $n$, they are always perpendicular to each other in canted AFMs, and a strong DMI vector $d$ enforces a fixed chirality between $m$ and $n$ as demonstrated above in cAFM $Mn_5Si_3$. Without losing the generality, we start by assuming $d$ to be perpendicular to the plane of $m$ and $n$, which forms an orthogonal coordinate system as presented in Fig. 3A. To achieve requirement (2), $n$ need to be driven into rotation along the $n \times p$ direction through the AFM exchange torque induced by the spin polarization $p$, which can be usually generated from the spin Hall effect of heavy metals such as Pt adjacent to $Mn_5Si_3$. We first consider $p$ perpendicular to the plane of $m$ and $n$, i.e., along $d$, and $n$ will roughly rotate circularly in this plane (Fig. 3A) for this scenario. However, due to equal transition barriers of $n_+$ to $n_-$ and $n_-$ to $n_+$ ($H = 0$ case of Fig. 3B), $n$ will rotate continuously driven by $p$, and finally non-deterministically relaxes to $n_+$ or $n_-$ after $p$ is withdrawn. To address this issue, we propose to construct asymmetric switching barriers through a fixed magnetic field $H$ acting on nonzero $m$ as shown in Fig. 3B ($H > 0$ case), and the asymmetry will be maximized when $H$ is perpendicular to $m$ as shown in Fig. 3A. In this case, through choosing a positive $p$ with suitable magnitude, $n$ can rotate from $n_+$ to $n_-$ through the lower barrier but cannot further climb over the higher barrier from $n_-$ to $n_+$, and thus stays at the intermediate state $n'_-$. When $p$ is withdrawn, $n$ will naturally stabilize as $n_-$. Similarly, by applying another $p$ with opposite direction, $n$ can switch from $n_-$ back to $n_+$ through the same lower energy barrier. Thus, the desired deterministic 180º switching between $n_+$ and $n_-$ can be electrically realized by damping-like SOT with a tiny external magnetic field.

This SOT mechanism with asymmetric energy barrier for 180º switching between $n_+$ and $n_-$ can further be confirmed explicitly by atomic spin simulations. Under a positive $p$, $n$ will rotate clockwise accompanied by *m* moving towards the hemisphere favored by positive *H*, and then settles at an intermediate state *n'*₋ (Fig. 3C). After SOT is turned off, *n* will naturally relax to the local energy minimum *n*₋ to achieve the deterministic 180º switching from *n*₊ to *n*₋ (fig. S10 for other perspectives). Similarly, *n* will switch back to *n*₊ from *n*₋ deterministically with a reversing *p* through anti-clockwise rotation (Fig. 3D). Reversing the direction of *H* also gives rise to deterministic 180º switching with reversing switching polarity (fig. S11). The same conclusion can be drawn through similar mechanisms for arbitrary orientations of *p* and *H* revealed by atomic spin simulations, even when *d* is not perpendicular to *m* and *n* (Supplementary text IV, fig. S12–13). Note that this mechanism is not restricted to $Mn_5Si_3$ but can be easily generalized to other AFMs with tiny intrinsic net moment.

**Experimental demonstration of electrical 180º switching of the Néel vector**

We carried out SOT switching experiments on $Mn_5Si_3$(0001)/Pt cross-bars of 6 μm × 4 μm (Fig. 4A) based on the mechanism above, where Pt serves as the spin source. Electrical pulses of different magnitude were introduced into the write channel $WI^+$–$WI^-$ with an assistant field *H* along the same direction, and the Hall voltage was collected simultaneously along $RI^+$–$RI^-$ to read the direction of *n* through AHE. As expected, 180º switching of *n* in cAFM $Mn_5Si_3$ are observed under various temperatures (fig. S14), and the results at 180 K are representatively presented in Fig. 4B. The largest switching ratio reaches 41%, and such an incomplete switching is most likely due to the existence of microscopical multidomain and local defects pinning (fig. S15). Opposite switching polarity can be observed if the polarity of current pulse or *H* is reversed, consistent with aforementioned theoretical results and matching basic symmetry requirements for deterministic SOT switching (*9, 10*). No switching can be observed with a high *H* of 8 kOe, which confines the direction of *m* and thus also confines the direction of *n*, further verifying that the change of Hall voltage is due to the switching of *n* instead of thermal artifacts. Moreover, AHE can also be observed in the temperature range of ncAFM $Mn_5Si_3$ (below 60 K), but electrical 180º switching cannot be achieved, further demonstrating it is *n* that can be switched electrically (fig. S16).

We now show that the 180º SOT switching of *n* in cAFM $Mn_5Si_3$ is fundamentally different from ferromagnets and ncAFMs such as $Mn_3Sn$ (*26, 27, 51-55*). In ferromagnets, the assistant field *H* can effectively assist the SOT to overcome MAE for switching, as characterized by a linear decrease of the critical switching current density $J_c$ with increasing *H* (*44*). In contrast, for cAFM $Mn_5Si_3$, *H* only eliminates the degeneracy of energy barriers between *n*₊ to *n*₋ and *n*₋ to *n*₊, but barely helps *n* to overcome MAE. That is because it is the large AFM exchange energy that determines the tilting of *n* towards *p* as a prerequisite to obtain nonzero AFM exchange torque for overcoming MAE, where AFM exchange energy is much larger than the Zeeman energy induced by *H* acting on the negligible *m*. Thus, $J_c$ should be very insensitive to *H* for cAFM $Mn_5Si_3$. As presented in Fig. 4C and fig. S17, we carried out SOT switching measurements under different assistant field *H*, and the experimentally-extracted $J_c$ (Supplementary text V) keeps almost unchanged with the increase of *H* for cAFM $Mn_5Si_3$. Besides, unlike uniaxial Néel vector *n* as order parameter perpendicular to *m* in cAFM $Mn_5Si_3$, cluster magnetic octuple as the order parameter with complex magnetic anisotropy in ncAFM $Mn_3Sn$ is parallel to its intrinsic *m*, leading to different dynamic behaviors under *p* and *H* of specific symmetry (*26, 27, 51-53*).

To investigate the efficiency for 180º switching of *n* in cAFM $Mn_5Si_3$, we calculate the ratio of AHE coercivity to critical switching current density $\mu_0 H_c/J_c$ as a figure of merit (*52*). For the averaged $J_c$ of 49 MA cm$^{-2}$ at 180 K (fig. S15), it turns out to be about 5 mT cm$^2$

MA$^{-1}$, which is 5 times higher than traditional ferromagnetic systems such as Pt/Co/AlO$_x$ of approximately 1 mT cm$^2$ MA$^{-1}$ (*52, 56*), indicating that the 180º SOT switching of *n* in cAFM Mn$_5$Si$_3$ shows high efficiency. Moreover, we test the endurance for 180º switching of *n* by applying electrical pulses with opposite directions alternately. The Hall voltage can be switched between high and low values corresponding to information states between "0" and "1" (Fig. 4D), which is reversible and reproducible. The device does not show degradation after five hundreds of cycles, manifesting its robustness. As a result, spin-splitting Mn$_5$Si$_3$ shows unique advantages to construct electrical-controllable AFM-RAM with low power-consumption, high reliability, and exceptional abundance in addition to its intrinsic high density and high speed.

**Discussion**

We achieve electrical 180º switching of the Néel vector in spin-splitting cAFM Mn$_5$Si$_3$. The demonstration of electrical 180º switching of the Néel vector involves two steps. One is to prove that it is the 180º switching of the Néel vector instead of net moment that tunes the spin-splitting band structure and brings about opposite anomalous Hall conductivity, which makes AHE as a unique readout method of 180º switching of Néel vector. It is supported by both systematic theoretical calculations and several control experiments including thickness-dependent and channel-width-dependent measurements. The other is to design SOT switching mechanism with asymmetric energy barriers between $n_+$ and $n_-$. It is experimentally implemented and confirmed by realizing current and assistant field polarity dependent 180º Néel vector switching, and further verified by assistant field magnitude dependent measurements and non-switching of ncAFM Mn$_5$Si$_3$. Based on electrical 180º switching and electrical readout of the Néel vector in Mn$_5$Si$_3$, electrical-controllable AFM devices with "0" and "1" resistance states were fabricated, which show high efficiency and high reproducibility. Our results establish the groundwork for practical applications of spin-splitting AFMs in next-generation information technology and serve as the basis of exploring other intriguing properties in spin-splitting AFMs arising from interactions among different degrees of freedom like spin, valley and charge.

**Materials and Methods**

**Sample preparation**

6 nm Mn$_5$Si$_3$(0001) films were grown on Al$_2$O$_3$(0001) substrates by co-sputtering of Mn and Si at 465 ºC with a rate of 0.4 Å s$^{-1}$ under a base pressure below 5×10$^{-8}$ Torr, followed by annealing at 600 ºC for 2 hours. After that, 10 nm Pt was deposited in-situ by magnetron sputtering at room temperature. 6 μm×4 μm cross bars for SOT switching measurements were patterned using optical lithography combined with Ar ion milling. Then Cr(10 nm)/Cu(70 nm) electrode was e-beam evaporated for wire bonding. For room temperature and low temperature X-ray diffraction (XRD) measurements, 80 nm Mn$_5$Si$_3$(0001) films were prepared to obtain diffraction peaks with substantial intensity for precisely determining the lattice parameter. Besides, 130 nm Mn$_5$Si$_3$(0001) films were grown for control measurements. For size-dependent measurements of the AHE, e-beam lithography was used to fabricate cross bars with width varying from 100 nm to 5 μm. All of the samples were kept in a glove box with O$_2$ and H$_2$O < 0.01 parts per million to prevent degradation or oxidation.

## Characterizations

Cross-sectional transmission electron microscopy (TEM) images for $Mn_5Si_3$(6 nm)/Pt(10 nm) were collected at room temperature using a commercial TEM system (JEM-2100F). Magnetic hysteresis curves were collected in commercial superconducting quantum interference device (SQUID, Quantum Design), where the diamagnetic contribution of $Al_2O_3$(0001) substrate was subtracted. Magnetization-temperature curves were also collected from SQUID without subtracting diamagnetic contribution of $Al_2O_3$(0001) substrate. $\theta$-$2\theta$ XRD measurement at room temperature was carried out at Rigaku Smartlab. In-situ $\theta$-$2\theta$ XRD at low temperatures were performed at BL02U2 Beamline from Shanghai Synchrotron Radiation Facility. Atomic force microscopy (AFM) images were collected at Bruker Dimension FastScan.

## Transport measurements

$\rho_{xx}$ and $\rho_{yx}$ of $Mn_5Si_3$ films were measured using a standard four-terminal method in commercial Physical Property Measurement System (PPMS, Quantum Design). Ordinary Hall resistivity that is linear to magnetic field was subtracted from $\rho_{yx}$. For SOT switching measurements, 1 ms writing pulses within the range of 46 mA were added in the $WI^+$-$WI^-$ channel with an assistant field $H$ applied along the $WI^+$-$WI^-$ channel, followed by waiting for 10 s before collecting the Hall voltage in the $RV^+$-$RV^-$ channel by a nanovoltmeter with a base current of 2 mA (Fig. 4A).

## Time-resolved reflectivity measurements

Time-resolved reflectivity measurements were carried out by a Ti:sapphire oscillator (center wavelength 800 nm, repetition rate 80 MHz, pulse width 20 fs) in a typical wavelength-degenerate pump-probe setup. Both the pump and probe beams were linearly polarized in a cross-polarization configuration to block the pump scattering. The pump beam was focused to 30 μm on the sample with a fixed fluence at 35.4 μJ/cm$^2$. The focal spot size and fluence of the probe beam were about 2 and 40 times smaller than those of the pump beam, respectively. The intensity of the pump beam was modulated by an optical chopper at 3 kHz to facilitate lock-in detection. The reflectivity signal was detected by a balanced detector to suppress laser power fluctuations. The sample was kept in a cryostat under vacuum of better than $1 \times 10^{-3}$ mbar during experiments.

## Magneto-optical Kerr effect measurements

The polar magneto-optical Kerr effect measurements were carried out using a power-stabilized 633 nm HeNe laser. After transmitting through a linear polarizer, the light (25 μW) was focused to a ~1 μm spot on the sample by a 40x reflective objective at normal incidence to avoid the large backgrounds that occur when a typical lens is used. The sample was mounted in the vacuum chamber of an optical superconducting magnet system with the magnetic field applied perpendicular to the sample plane (Faraday geometry). The reflected beam was modulated at ~50 kHz by a photo-elastic modulator, split by a Wollaston prism, and detected using a balanced photodiode with a standard optical bridge arrangement. The resulting 50 and 100 kHz modulations detected by lock-in amplifiers then correspond to the ellipticity and rotation angle of the beam respectively. We additionally modulated the intensity of the beam with a lower-frequency (~1319 Hz) chopper to measure the DC signal for normalization using a third lock-in amplifier. The background from the optic window and a linear background were subtracted.

## First-principles calculations

First-principles calculations were performed based on density functional theory (57) (DFT) as implemented in Vienna ab initio simulation package (VASP) (58). Exchange-correlation interaction was described by the Perdew-Burke-Ernzerhof (PBE) parametrization of generalized gradient approximation (GGA) (59). We also used Perdew-Wang (PW91) (60) and local density approximation (LDA) exchange-correlation functional, where similar results were obtained. Structures were relaxed until the force on each atom is less than 0.01 eV/Å. The cutoff energy and electronic iteration convergence criterion were set to 400 eV and $10^{-5}$ eV, respectively. To model the Brillouin zone, a Monkhorst-Pack (MP) $k$-grid mesh (61) of $7 \times 7 \times 9$ was used. For MAE calculations, self-consistent charge density was read to get a converged result. Four-states method was utilized to evaluate the strength of DMI. Berry curvature and AHC were calculated based on ab initio tight-binding models with all parameters extracted from first-principles calculations using the FPLO software (62).

**Atomic spin simulations of magnetic hysteresis**

We carried out atomic spin simulations on VAMPIRE software (63), which can define specific lattice parameters $a$ and $c$ consistent with our hexagonal cAFM phase $Mn_5Si_3$ thin film of 6.902 Å and 4.795 Å, respectively. Discretized $50 \times 50 \times 10$ units cells in the $x$-$y$-$z$ direction were used to model the $Mn_5Si_3$ thin film. The Hamiltonian of the system at 0 K can be written as following:

$$H_{\text{total}} = \sum_{<ip,jq>} J_{ip,jq} m_{ip} \cdot m_{jq} + \sum_{<ip,jq>} d_{ip,jq} \cdot (m_{ip} \times m_{jq}) - K \sum_{ip} (k \cdot m_{ip})^2 + H_Z$$

Here, $i$ and $j$ denote unit cell, $p$ and $q$ represent sublattice ($p, q = 1, 2, 3, 4$). $H_{\text{total}}$ is the total Hamiltonian and $m_{ip}$ ($m_{jq}$) is the unit magnetic moment at sublattice $Mn_p$ ($Mn_q$) of unit cell $i$ ($j$) with the magnitude of 2.4 $\mu_B$. $J_{ip,jq}$, $d_{ip,jq}$, and $K$ are the exchange interaction constant, the DMI vector, and the uniaxial anisotropy constant, respectively. $H_Z$ is the Zeeman energy under magnetic field, equaling $-\mu_0 m_s \sum_{ip} m_{ip} \cdot H$, where $\mu_0$, $m_s$, and $H$ are the vacuum permeability, the saturation magnetization of magnetic Mn atom, and the external field, respectively. $k$ is the direction of easy axis, which was set to be (1,1,1) direction in the cartesian coordinate system with the uniaxial anisotropy constant $K$ of 0.1 meV from DFT estimations (Supplementary text III). Four exchange interactions were considered, namely $J_1 = -12.23$ meV, $J_2 = -2.16$ meV, $J_3 = 3.98$ meV, and $J_4 = 2.89$ meV (fig. S8A) with magnitude consistent with former DFT calculation results (64, 65). Note that $J_4$ interaction was set to be ferromagnetic to eliminate the $t\mathcal{T}$ symmetry for consistency with experimentally measured non-zero $\sigma_{xy}$ (Supplementary text III). The DMI vector was set along $x$ axis with a magnitude of 1.35 meV from DFT calculations and the experimentally measured out-of-plane AHE-related net moment $m$ (Supplementary text II). These settings of basic material parameters give a simulated magnetic hysteresis that is analogous with SQUID-derived $m$ under 150 K (Fig. 2C, Supplementary text II), illustrating that these parameters are suitable for modeling the magnetic switching behavior of cAFM $Mn_5Si_3$. Specifically speaking, the Landau-lifshitz-Gilbert (LLG) equation for each magnetic Mn atom is written as:

$$\frac{dm_{ip}}{dt} = -|\gamma| m_{ip} \times H_{\text{eff},ip} + \alpha m_{ip} \times \frac{dm_{ip}}{dt}$$

Where $H_{\text{eff},ip}$ is the effective magnetic field acting on the magnetic Mn atom at sublattice $p$ of unit cell $i$. $H_{\text{eff},ip} = -\frac{\partial H_{\text{total}}}{\partial m_{ip}} + H_{\text{th},ip}$, where $H_{\text{th},ip}$ is the Gaussian stochastic magnetic field to describe the effect of finite temperature (66). $\gamma$ and $\alpha$ are the gyromagnetic ratio of the electron and the Gilbert damping coefficient, equaling 176 s$^{-1}$ T$^{-1}$ and 0.1, respectively.

A fine iteration time step $\Delta t$ of 0.1 fs was used to precisely capture the dynamics of atomic magnetic moments. First, the system was initialized with random orientations of atomic magnetic moments, and then relaxed into a perfect G-type AFM ordering at 0 K (fig. S8B). After applying negative magnetic field, each atomic magnetic moment can be switched by 180º (fig. S8C). The simulated $m$ and $n$ for each unit cell should be defined as $m = (m_1 + m_2 + m_3 + m_4) / 4$ and $n = (m_1 - m_2 + m_3 - m_4) / 4$, which further equaling $(m_1 + m_2) / 2$ and $(m_1 - m_2) / 2$ since $m_1 = m_3$ and $m_2 = m_4$ due to $J_3$. Thus, the 180º switching of each atomic magnetic moment is consistent with the 180º switching of $m$ and $n$. Then the hysteresis was simulated at 150 K (Fig. 2D), and 180º switching of $m$ and $n$ can also be observed (Fig. 2E), where $m$ and $n$ on all of unit cells were averaged.

**Atomic spin simulations of SOT switching**

To understand the mechanism for SOT 180º switching of the Néel vector $n$ of cAFM $Mn_5Si_3$, we consider the damping-like effective field $H_{DL}$ of the spin polarization $p$ generated by the spin Hall effect of Pt. Then the LLG equation under SOT can be written as:

$$\frac{dm_{ip}}{dt} = -|\gamma|m_{ip} \times H_{\text{eff},ip} + \alpha m_{ip} \times \frac{dm_{ip}}{dt} + \xi_{DL} m_{ip} \times (p \times m_{ip})$$

$$= -|\gamma|m_{ip} \times H_{\text{eff},ip} + \alpha m_{ip} \times \frac{dm_{ip}}{dt} + |\gamma|m_{ip} \times H_{DL,ip}$$

Where $H_{DL,ip}$ is the damping-like effective field of SOT acting on the magnetic Mn atom at sublattice $p$ ($p = 1, 2, 3, 4$) of unit cell $i$. $p$ is the unit vector for the spin polarization. $\xi_{DL}$ is the magnitude of damping-like SOT, which has linear relationship with the magnitude of $H_{DL,ip}$ and current density $J$ (Supplementary text III). Note that for the simulation of SOT switching, 5×5×5 units cells were used to save calculation time, which gives consistent results with that of 50×50×10 units cells. In analogous to the simulations for magnetic hysteresis, the system was initialized with random orientations of atomic magnetic moments, and relaxed into G-type AFM ordering before adding a 0.1 ns long electrical pulse with transverse $p$ under assistant field $H$. The magnitude of assistant field $H$ was fixed at 10 kOe. This magnitude of $H$ is reasonable, because the simulations of SOT switching were carried out at 0 K, and the coercive field of simulated magnetic hysteresis at 0 K is around 150 kOe. It means that $H$ of no more than 10% of the coercive field can assist the deterministic 180º switching of $n$, consistent with experimental results. $m$ and $n$ on all of unit cells were averaged to plot trajectories of $m$ and $n$ under $p$. The settings of pulse lengths and random seeds does not influence simulation results, proving their reliability (Supplementary text VI). The influence of field-like spin-orbit torques was also investigated by adding field-like effective field $H_{FL}$ as a term of $|\gamma|m_{ip} \times H_{FL,ip}$. It turns out that $H_{FL}$ has a negative influence on SOT switching, which increases the critical switching current density (Supplementary text VII).

**Acknowledgments:**
We acknowledge BL02U2 of Shanghai Synchrotron Radiation Facility. Some devices were fabricated via an Ultraviolet Maskless Lithography machine (Model: UV Litho-ACA, TuoTuo Technology).

**Funding:**
National Key R&D Program of China (Grant No. 2022YFA1402603, 2021YFB3601301, 2021YFA1401500, 2020YFA0308800, and 2021YFA1400100)
National Natural Science Foundation of China (Grant No. 52225106, 12241404, 12074212, and 12022416)
Hong Kong Research Grants Council (Grant No. 16304523, 16303821, 16306722)
Natural Science Foundation of Beijing, China (Grant No. JQ20010)

**Author contributions:**
Conceptualization: L. H., X. F., Xin. C., J. D., L. L., Yi. Z., W. Z., L. Y., C. S., J. L., F. P., Data curation: X. F., Xin. C., L. L., Y. L., Yo. Z., Q. W., C. S., J. L., F. P., Formal analysis: L. H., X. F., R. P., Xin. C., J. D., L. L., Y. L., Ya. Z., C. S., J. L., F. P., Funding acquisition: Xin. C., L. Y., C. S., J. L., F. P., Investigation: L. H., X. F., J. D., L. L., Y. L., S. L., Ya. Z., C. S., J. L., F. P., Methodology: L. H., X. F., Xin. C., J. D., L. L., C. C., L. Y., C. S., J. L., F. P., Project administration: L. H., L. Y., C. S., J. L., F. P., Resources: J. D., Yo. Z., Q. W., L. Y., C. S., J. L., F. P., Software: L. H., X. F., Xin. C., Ya. Z., C. S., J. L., F. P., Supervision: L. H., Q. W., L. Y., C. S., J. L., F. P., Validation: L. H., X. F., R. P., J. D., L. L., Y. L., Q. W., Xia. C., L. Y., Ya. Z., C. S., J. L., F. P., Visualization: L. H., X. F., Xin. C., J. D., Y. L., W. Z., H. B., Yo. Z., C. S., J. L., F. P., Writing-original draft: L. H., X. F., Xin. C., J. D., Y. L., Y. L., Yo. Z., C. C., C. S., J. L., F. P., Writing-review and editing: L. H., X. F., Xin. C., J. D., L. L., Y. L., W. Z., Yo. Z., S. L., L. Y., C. S., J. L., F. P.

**Competing interests:**
Authors declare that they have no competing interests.

**Data and materials availability:**
All data needed to evaluate the conclusions in the paper are present in the paper and/or the Supplementary Materials.

## Supplementary Materials

Supplementary Text
Figs. S1 to S20
Movies S1
References 57-75

## Figures and Tables

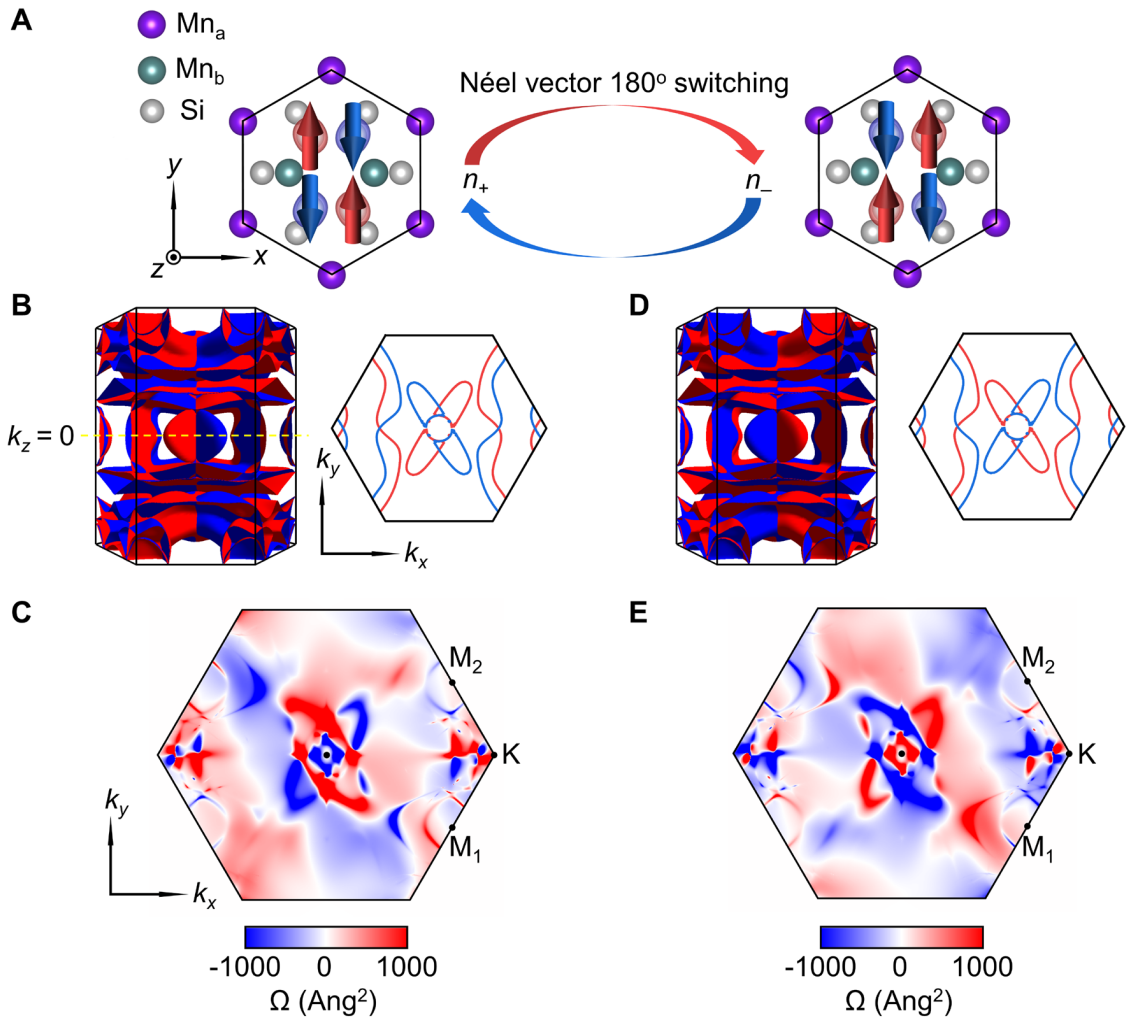

**Fig. 1. Spin-splitting band and Berry curvature in Mn$_5$Si$_3$ of opposite Néel vectors.** (**A**) Crystal and magnetic structure of G-type cAFM Mn$_5$Si$_3$ with opposite magnetization density represented by red and blue isosurfaces. (**B**, **D**) Fermi surface and its contour plot in the $k_z = 0$ plane. (**C**, **E**) Berry curvature in the $k_z = 0$ plane. When the Néel vector is 180°-switched, both spin-splitting band and Berry curvature are flipped.

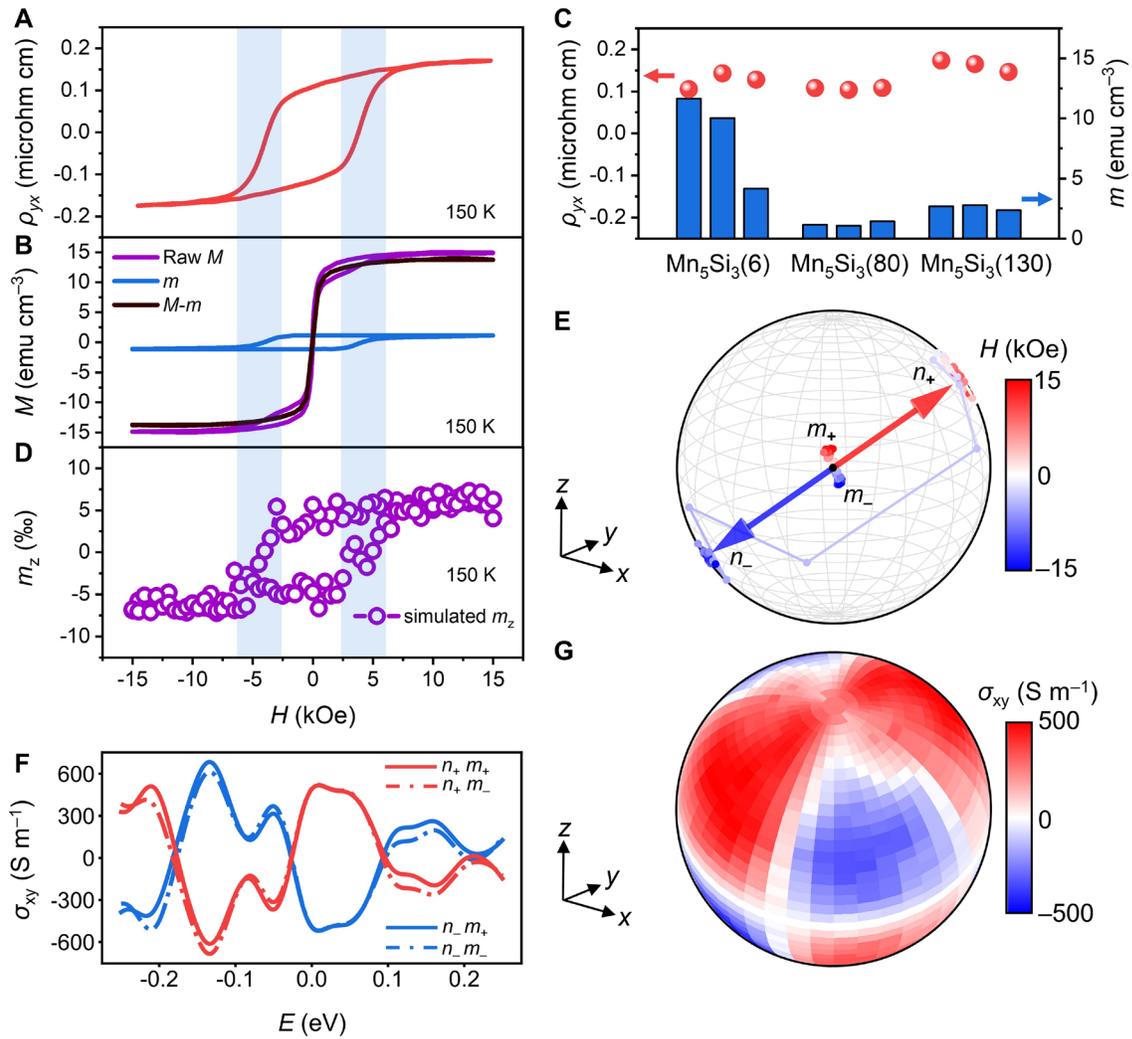

**Fig. 2. Electrical detection of the Néel vector of Mn$_5$Si$_3$ by the anomalous Hall effect.** (**A**) Hysteresis of Hall resistivity $\rho_{yx}$ and (**B**) magnetization $M$ under out-of-plane magnetic field at 150 K, within the cAFM phase of Mn$_5$Si$_3$. $M$ can be decomposed into DMI-induced $m$ as blue line and defect-induced $M$-$m$ as black line. (**C**) Comparison between $\rho_{yx}$ and $m$ of Mn$_5$Si$_3$ thin films with different thickness at 150 K. The numbers in parentheses indicate the thickness of the film in nanometer. (**D**) Simulated hysteresis of out-of-plane net moment $m_z$ under out-of-plane magnetic field at 150 K. (**E**) Simulated trajectories of 180° switching of Néel vector $n$ and net moment $m$ by out-of-plane magnetic field from 15 kOe to –15 kOe at 150 K. $m$ and $n$ are defined based on the AFM sublattice of Mn$_5$Si$_3$ and averaged over all of simulation cells (Materials and Methods). (**F**) Calculated $\sigma_{xy}$ near the Fermi level for the Néel vector along $n_+$ or $n_-$ with net moment along $m_+$ or $m_-$ at 150 K. (**G**) Calculated $\sigma_{xy}$ at the Fermi level for $n$ along any direction at 150 K.

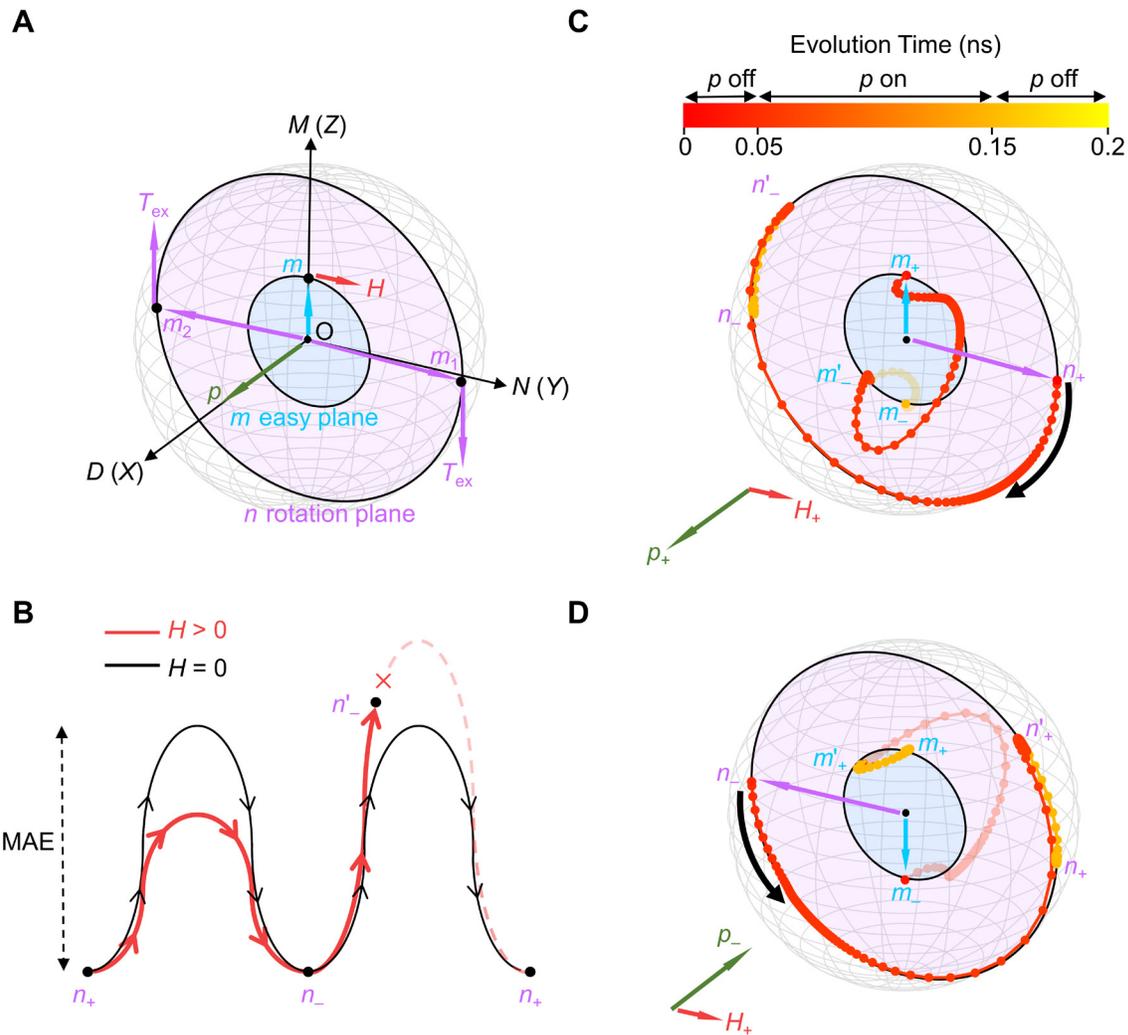

**Fig. 3. Mechanism for 180° switching of the Néel vector by SOT.** (**A**) Schematic of $n$ rotation driven by exchange torques $T_{ex}$ with $p$ along $+X$ and $n$ along $+Y$. $n$ can be simplified as $(m_1-m_2)/2$ for cAFM $Mn_5Si_3$ with four sublattices where $m_1 = m_3$ and $m_2 = m_4$ (Materials and Methods). Damping-like SOT shares the same direction for $m_1$ and $m_2$ on opposite AFM sublattice as $T_{DL} \sim m_1 \times (p \times m_1) \sim m_2 \times (p \times m_2)$. Hence, $T_{DL}$ pulls $m_1$ and $m_2$ towards the direction of $p$, bringing about exchange torques $T_{ex} \sim m_1 \times H_{ex} \sim m_2 \times H_{ex}$ due to the AFM exchange field $H_{ex}$, which alters on $m_1$ and $m_2$ to drive both $m_1$ and $m_2$ to rotate coherently. The rotation of $n$ is accompanied by the motion of $m$, which is favorite by magnetic field $H$ along $+Y$. (**B**) Schematic of energy barriers for the transition from $n_+$ to $n_-$ and $n_-$ to $n_+$. Without $H$, these two energy barriers are the same, determined by MAE. With non-zero $H$ such as $H > 0$, the degeneracy between them cannot be maintained. (**C**) Simulated switching trajectories of $m$ and $n$ for positive $p_+$ and positive $H_+$ as well as (**D**) negative $p_-$ and positive $H_+$. $m$ is magnified 10 times for better visualization. The color of trajectories indicates the evolution time, where $p$ is added from 0.05 ns to 0.15 ns.

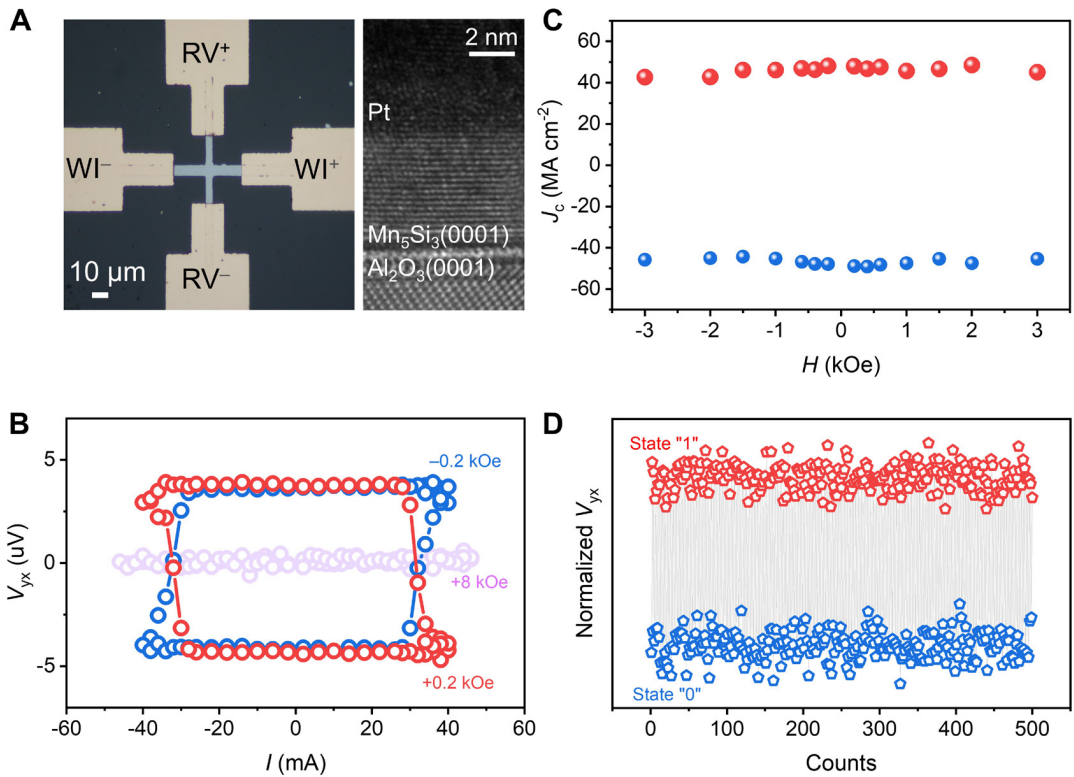

**Fig. 4. Electrical 180º switching of the Néel vector of Mn$_5$Si$_3$.** (**A**) Optical microscopy and transmission electron microscopy image of Mn$_5$Si$_3$(0001)/Pt cross-bars for electrical switching. WI$^+$ and WI$^-$ denote the write channel where the electrical pulse $I$ is added and RV$^+$ and RV$^-$ denote the read channel where the Hall voltage $V_{yx}$ is collected. (**B**) Hall voltage $V_{yx}$ as a function of electrical pulse $I$ for different $H$ of –0.2, 0.2, and 8 kOe at 180 K. (**C**) Critical switching current density $J_c$ under different assistant field $H$ at 180 K. (**D**) Continuous cycling between high $V_{yx}$ (state "1") and low $V_{yx}$ (state "0") by 180º switching of $n$ for five hundreds of electrical pulses at 180 K.